\documentclass[aps,prl,superscriptaddress,twocolumn]{revtex4-1}

\usepackage{graphicx}

\usepackage[usenames,dvipsnames]{color}

\begin{document}

\title{Dynamo Action in a Quasi-Keplerian Taylor-Couette Flow}

\author{Anna Guseva}
\email{anna.guseva@zarm.uni-bremen.de}
\affiliation{University of Bremen, Center of Applied Space Technology and Microgravity (ZARM),
28359 Bremen, Germany}
\author{Rainer Hollerbach}
\email{rh@maths.leeds.ac.uk}
\affiliation{School of Mathematics, University of Leeds, Leeds LS2 9JT, UK}
\author{Ashley P. Willis}
\affiliation{School of Mathematics and Statistics, University of Sheffield,
Sheffield S3 7RH, UK}
\author{Marc Avila$^1$}

\date{\today}

\begin{abstract}
We numerically compute the flow of an electrically conducting fluid in a Taylor-Couette geometry where the rotation rates of the inner and outer cylinders satisfy $\Omega_o/\Omega_i=(r_o/r_i)^{-3/2}$. In this quasi-Keplerian regime a non-magnetic system would be Rayleigh-stable for all Reynolds numbers $Re$, and the resulting purely azimuthal flow incapable of kinematic dynamo action for all magnetic Reynolds numbers $Rm$. For $Re=10^4$ and $Rm=10^5$ we demonstrate the existence of a finite-amplitude dynamo, whereby a suitable initial condition yields mutually sustaining turbulence and magnetic fields, even though neither could exist without the other. This dynamo solution results in significantly increased outward angular momentum transport, with the bulk of the transport being by Maxwell rather than Reynolds stresses.
\end{abstract}

\maketitle

The magnetic fields of planets, stars and entire galaxies are created by dynamo action, in which the motion of electrically conducting fluid stretches and thereby amplifies some original seed field. The details vary widely for different objects \cite{Roberts,Charbonneau,Widrow,Brandenburg1}, but it is generally believed that most sufficiently complicated fluid flows can act as dynamos, at least if the electrical conductivity is large enough. The onset of dynamo action then becomes a linear instability problem, with the electrical conductivity incorporated into the so-called magnetic Reynolds number $Rm$ as the control parameter. Once $Rm$ exceeds some critical value any infinitesimal seed field will grow exponentially in time. This process continues until the field is so strong that its associated Lorentz force alters the original flow and eventually stops further field amplification.

One astrophysical category where this process may not work quite so simply are accretion disks. The difficulty is that a Keplerian angular rotation profile, $\Omega(r)\sim r^{-3/2}$, fails the requirement to be `sufficiently complicated'. A flow consisting of only the single component $U_\phi=\Omega\,r$ is so simple that it will never yield dynamo action, no matter how large $Rm$ is taken to be. To explain the magnetic fields of accretion disks, the flow must therefore be more complicated than just the large-scale Keplerian rotation profile. The generally accepted explanation is that there is also small-scale turbulence \cite{Balbus1}. This naturally raises the question regarding the origin of this turbulence, especially since the familiar Rayleigh criterion \cite{Rayleigh} states that flows with angular momentum $\Omega r^2$ increasing outward are hydrodynamically stable. The Rayleigh criterion is admittedly a purely linear result, and thus does not exclude the possibility of a nonlinear, finite amplitude instability. Nevertheless, there are both experimental \cite{Ji1,Edlund} and numerical \cite{Lesur,Avila,Ostilla,Lopez,Shi} results which suggest that Keplerian rotation profiles are indeed stable even with respect to finite amplitude perturbations.

There is actually an easy way to bypass the Rayleigh criterion, namely by including magnetic fields. This leads to the magnetorotational instability (MRI), first discovered in the Taylor-Couette context in 1959 \cite{Velikhov}, and applied to accretion disks in 1991 \cite{Balbus2}. By using the tension in magnetic field lines to transfer angular momentum between fluid parcels, a key ingredient in the derivation of the Rayleigh criterion is invalidated, namely that without magnetic fields angular momentum is conserved not only globally but also locally on individual fluid parcels. The result is that in the presence of magnetic fields it is only the angular velocity $\Omega$ rather than the angular momentum $\Omega r^2$ which needs to be outwardly decreasing for the flow to be unstable. Keplerian profiles $\Omega\sim r^{-3/2}$ are thus Rayleigh-stable but MRI-unstable. Since its rediscovery in the astrophysical context, there has been enormous further interest in the MRI \cite{Balbus1}, including also the possibility of obtaining it and variants of it experimentally \cite{Rudiger,Ji2,Hollerbach1,Stefani,Flanagan}.

There is of course one remaining difficulty before the MRI can be invoked to explain the magnetic fields of accretion disks, in that it essentially leads to a `chicken and egg' type situation. If a (sufficiently strong) magnetic field were present, the disk would almost invariably be turbulent, via the MRI, which would likely yield a sufficiently complicated flow to act as a dynamo. However, before the dynamo is operating, where does the initial magnetic field come from? One possibility is that the entire dynamo process is a finite amplitude rather than a linear instability (as encountered also in other contexts, e.g.\ \cite{Christensen,Ponty,Krstulovic}). That is, an infinitesimally small seed field would yield neither turbulence nor a dynamo, but some sufficiently strong initial field could yield a configuration that permanently maintains both the turbulence and an associated magnetic field. The possibility of a dynamo of this type has been explored very extensively in local shearing-box simulations \cite{Brandenburg2,Hawley,Fromang,Yousef,Johansen,Riols,Kunz,Nauman}, but not in global calculations. In this \emph{Letter} we provide numerical evidence for the existence of such a finite amplitude dynamo in a global Taylor-Couette geometry, with the inner and outer cylinder's rotation rates set to be in the Rayleigh-stable regime.

We start with a Taylor-Couette system having nondimensional inner and outer cylinder radii $r_i=1$ and $r_o=2$. Periodicity is imposed in the axial direction, with length $L_z=1.4$. This periodicity in the axial direction is the most obvious difference between Taylor-Couette flows and accretion disks. The rotation rates of the two cylinders are fixed to satisfy $\Omega_o/\Omega_i=(r_o/r_i)^{-3/2}=0.35$, thereby matching $\Omega\sim r^{-3/2}$ at the boundaries to constitute what is known as a quasi-Keplerian system. In particular, the resulting basic state flow profile is Rayleigh-stable in the purely hydrodynamic regime. By contrast, previous numerical Taylor-Couette dynamos \cite{Willis,Nore1,Gissinger,Nore2} have been in the regime where the flow is already hydrodynamically unstable, and is thus sufficiently complicated to work even as a kinematic dynamo. There have also been two liquid sodium dynamo experiments \cite{Gailitis,Monchaux} in cylindrical geometry, but again in a regime that does not rely on finite amplitude instabilities.

The governing equations for the fluid flow $\bf U$ and the magnetic field $\bf B$ are
$$\frac{\partial{\bf U}}{\partial t} + {\bf U\cdot\nabla U}=-\frac{1}{\rho}\,\nabla p
 + \nu\nabla^2{\bf U} + \frac{1}{\mu_0\rho}\,{\bf(\nabla\times B)\times B},$$
$$\frac{\partial{\bf B}}{\partial t}=\eta\nabla^2{\bf B} + \nabla\times({\bf U\times B}),$$
together with $\nabla\cdot{\bf U}=\nabla\cdot{\bf B}=0$. The associated boundary conditions at both cylinders are no-slip for $\bf U$ and insulating for $\bf B$. Here $p$ is the pressure, $\rho$ is the density, $\mu_0$ the permeability, $\nu$ the viscosity, and $\eta$ the magnetic diffusivity (inversely proportional to the electrical conductivity).

We nondimensionalize length by $r_i$, time by $\Omega_i^{-1}$, $\bf U$ by $\Omega_i r_i$, and $\bf B$ by $\Omega_i r_i \sqrt{\rho\mu_0}$. The relevant nondimensional parameters measuring the rotation rates are the ordinary and magnetic Reynolds numbers
$$Re=\frac{\Omega_i r_i^2}{\nu},\qquad Rm=\frac{\Omega_i r_i^2}{\eta}.$$
The ratio $Rm/Re=\nu/\eta$ is a material property of the fluid, known as the magnetic Prandtl number $Pm$. Liquid metals all have $Pm\le O(10^{-5})$, but astrophysical plasmas can have a much greater range, including $Pm\ge O(1)$.

In previous work \cite{Guseva1,Guseva2,Guseva3} we considered turbulent Taylor-Couette flows in the presence of an externally imposed azimuthal magnetic field $B_0(r_i/r)\,{\bf\hat e}_\phi$, which guarantees the existence of an instability, the so-called azimuthal magnetorotational instability \cite{Hollerbach2,Seilmayer}. As our finite amplitude initial condition here, we took a turbulent solution from this work, with $Re=Rm=10^4$. The external field was then switched off, and two separate runs were performed. For the first the two Reynolds numbers were both kept at $10^4$; for the second the magnetic Reynolds number was increased to $10^5$. The resolution for the first run was 280 points in radial direction, 600 azimuthal and 400 axial Fourier modes. For the second run the resolution was 480 points, 1024 azimuthal and 512 axial modes. For a full description of the numerical code see \cite{Guseva1}.

\begin{figure}
 \begin{center}
   \includegraphics[width=0.97\linewidth]{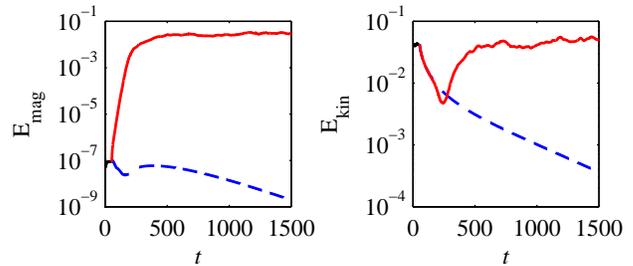}
 \end{center}
\caption{Magnetic (left) and kinetic (right) energies of the runs at $Rm=10^4$ (dashed blue lines) and $Rm=10^5$ (solid red lines), as functions of time.  The short line segments for $t\le50$ correspond to the initial condition before the external field is switched off \cite{Guseva2}. For the kinetic energies the laminar profile has been subtracted from $\bf U$, to better illustrate the relaminarisation of the $Pm=1$ run.} 
\label{energies}
\end{figure}

Figure \ref{energies} shows how the magnetic and kinetic energies for the two runs evolve in time. The $Rm=10^4$ case is clearly not a dynamo -- after some minor initial transients the magnetic energy starts to decay, while $\bf U$ relaxes back toward the basic quasi-Keplerian profile.  By contrast, the $Rm=10^5$ case shows an immediate dramatic increase in the magnetic energy, followed by saturation to a statistically steady state.  After some quite substantial transient adjustments, the flow also equilibrates to a statistically steady state.

\begin{figure*}
 \begin{center}
  \includegraphics[width=0.95\linewidth]{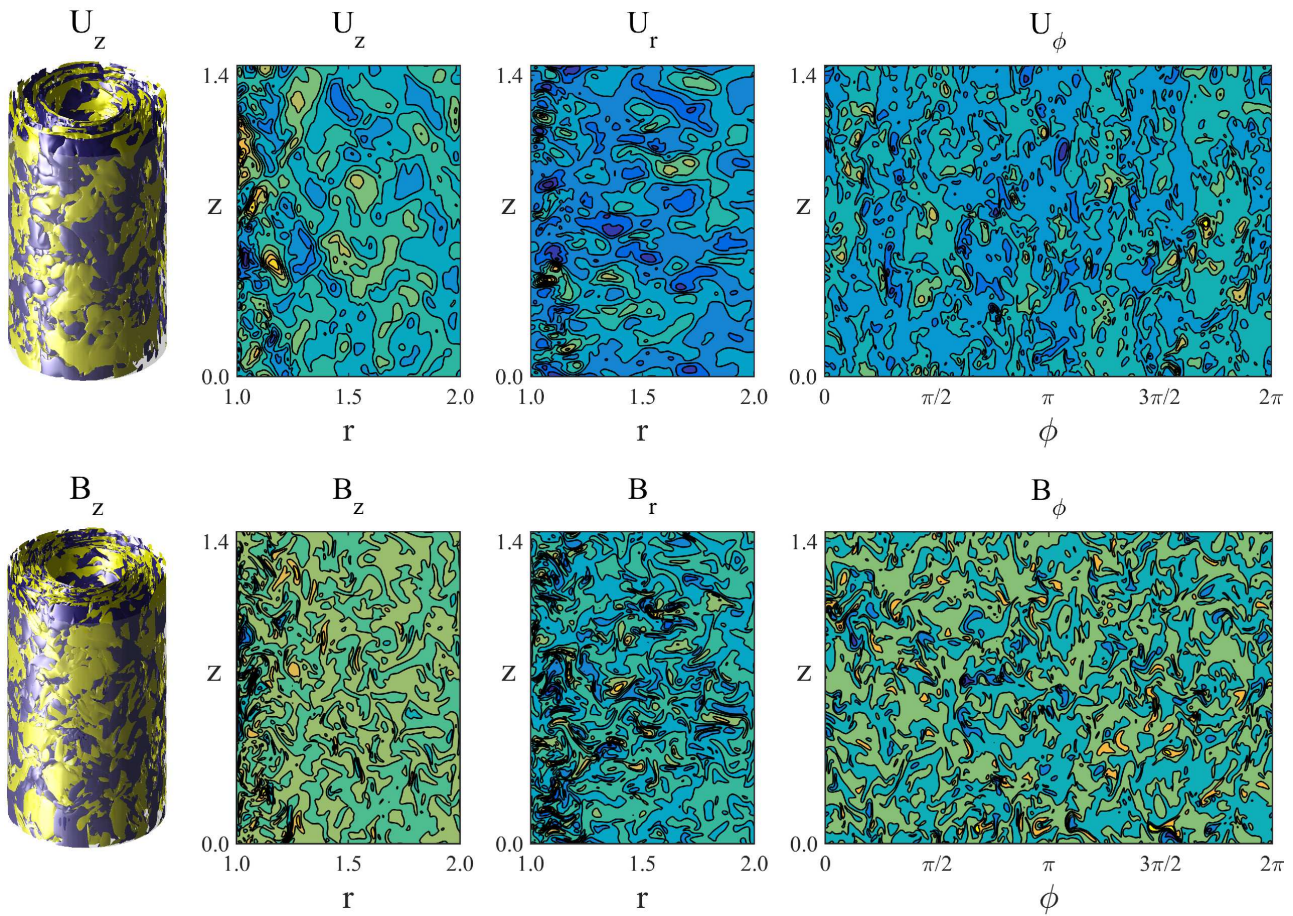}
 \end{center}
\caption{The top row shows different views of $\bf U$; from left to right the four panels show isosurfaces of $U_z=\pm0.01$, contours of $U_z$ and $U_r$ on slices in the $(r,z)$ meridional plane, and contours of $U_\phi$ on the unrolled cylinder at mid-gap $(r=1.5)$. The bottom row shows equivalent snapshots of $\bf B$.} 
 \label{snapshots}
\end{figure*}

\begin{figure}
 \begin{center}
  \includegraphics[width=0.8\linewidth]{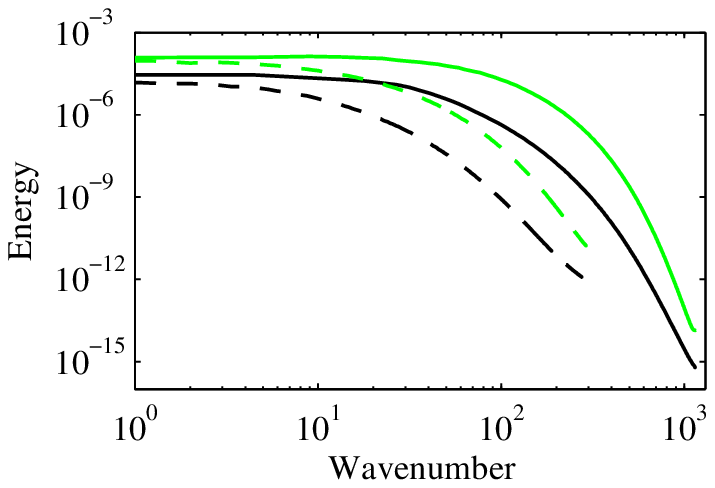}
 \end{center}
\caption{Time-averaged kinetic (black) and magnetic (green) energy spectra versus axial (solid) and azimuthal (dashed) wavenumber for the $Rm=10^5$ dynamo case. These spectra are computed at mid-gap $(r=1.5)$ only, but spectra averaged over the entire volume are qualitatively similar.}
\label{spectra}
\end{figure}

Figure \ref{snapshots} shows snapshots of $\bf U$ and $\bf B$ for the $Rm=10^5$ dynamo.  Both are seen to exhibit strong small-scale structures in all three directions.  The meridional slices indicate a certain concentration of structures toward the inner boundary, but otherwise no clearly discernible boundary layer structure.  As one would expect based on these snapshots, the energy spectra are not concentrated at the largest scales, but instead contain substantial energy out to quite large wavenumbers.  As shown in Figure \ref{spectra}, the spectra in both $z$ and $\phi$ are almost flat over one or even two orders of magnitude in wavenumber before dropping off.

Closely related to these spectra are the lengthscales
$$l_U^2=\frac{\int{\bf U}^2\,dV}{\int(\nabla\times{\bf U})^2\,dV},\quad
  l_B^2=\frac{\int{\bf B}^2\,dV}{\int(\nabla\times{\bf B})^2\,dV},$$
where the integrals are over the entire volume. The  instantaneous values from the end of the run give $l_U=4.5\cdot10^{-2}$ and $l_B=9.8\cdot10^{-3}$, broadly consistent with the spectra in Figure \ref{spectra}, as well as the $O(Rm^{-1/2})$ lengthscale on which a small-scale dynamo would be expected to operate \cite{Dormy}.  Note also that the diffusive timescales corresponding to these lengthscales, $Re\cdot l_U^2\approx20$ and $Rm\cdot l_B^2\approx10$, are both very short compared with the $t=1500$ integration time in Figure \ref{energies}, providing further evidence that this is indeed a permanent dynamo and not just remaining transients.

\begin{figure}
 \begin{center}
  \includegraphics[width=0.8\linewidth]{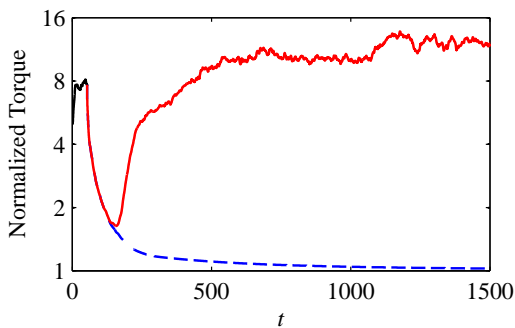}
\end{center}
\caption{\small Normalized torque as a function of time for $Rm=10^4$ (dashed) and $Rm=10^5$ (solid). The short black segment for $t\le50$ again corresponds to the initial condition.}
\label{torque}
\end{figure}

A quantity of particular interest in the accretion disk context is the associated outward angular momentum transport, since it is this which determines the rate at which matter actually accretes onto the central object. In the Taylor-Couette context this angular momentum transport is very easily quantified simply by considering the torques on the inner and outer cylinders (where the time-averaged torques are necessarily equal and opposite in a statistically steady state). Figure \ref{torque} shows how the normalized torque (that is, scaled by its value for the non-magnetic laminar $U_\phi$-only profile) evolves in time. For $Rm=10^4$ it very quickly tends to one, consistent with the result in Figure \ref{energies} that $\bf U$ simply relaxes back toward the basic state. By contrast, for $Rm=10^5$ there are again substantial initial transients, but eventually the torque settles in to $O(10)$ times greater than the laminar value.

There are two other points worth mentioning in comparison with Figure \ref{energies}. First, whereas in Figure \ref{energies} the kinetic energy ultimately settled in to a value quite similar to the initial condition with the imposed field, the torque settles in to values around twice as large as for the initial condition. Second, after seemingly settling in, the torque exhibits fluctuations of similar intensity to either the kinetic or magnetic energies, if values are normalized about the maximum.

\begin{figure}
 \begin{center}
   \includegraphics[width=0.97\linewidth]{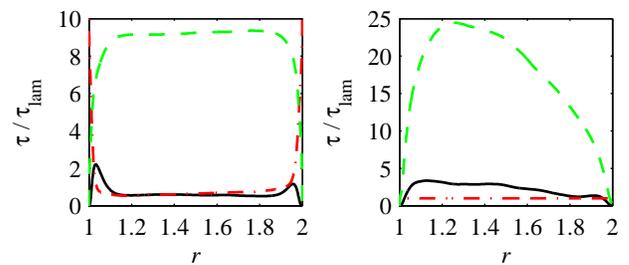}
 \end{center}
\caption{Both panels show the time-averaged stresses as functions of radius, with the Maxwell stress green-dashed, Reynolds black-solid, and viscous red-dash-dotted. The left panel is for the true Taylor-Couette system, where the $U_\phi$ profile is allowed to adjust as in figure \ref{profiles}. The right panel is for the system where the $U_\phi$ profile is forced to remain quasi-Keplerian throughout the interior.}
\label{stresses}
\end{figure}

These torque results already indicate that the turbulent, magnetic state at $Rm=10^{5}$ is very effective at angular momentum transport. A way of further quantifying this is shown in Figure \ref{stresses}a, where we compute the Maxwell, Reynolds and viscous stresses. We see that throughout the bulk of the interior the Maxwell stress dominates, with the Reynolds stress accounting for only around 10\%. This agrees with the point noted earlier that the magnetorotational instability works precisely by harnessing the tension in magnetic field lines to transport angular momentum.

\begin{figure}
 \begin{center}
  \includegraphics[width=0.97\linewidth]{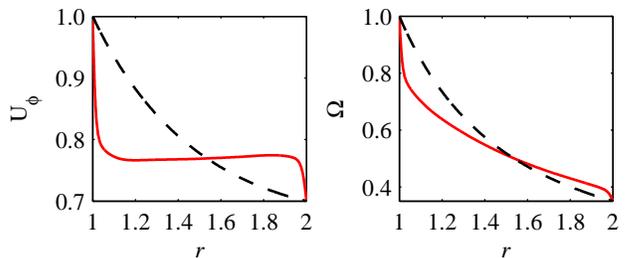}
 \end{center}
\caption{The left panel shows the time-averaged $U_\phi$ radial profile for the dynamo at $Rm=10^{5}$ (red solid), in comparison with the laminar quasi-Keplerian profile (black dashed). The right panel shows the corresponding angular velocities $\Omega=U_\phi/r$.}
\label{profiles}
\end{figure}

Next, Figure \ref{profiles} shows how the presence of turbulence modifies the time-averaged $U_\phi$ profile, and how it compares with the original quasi-Keplerian profile. The presence of boundary layers is as expected; since the torque right at the boundaries is purely viscous, if the torques in Figure \ref{torque} are $\sim10$ greater than the laminar values, then the gradients $\frac{d}{dr}\Omega$ at the boundaries must also be greater by the same amount. Note also how the viscous stresses in Figure \ref{stresses}a dominate within the boundary layers. It is not entirely clear though why the solution adjusts to have $U_\phi$ so uniform in the interior, as opposed to the angular velocity $\Omega$ (or some intermediate quantity $U_\phi/r^q$, with $0<q<1$). See also \cite{Brauckmann} who explore related issues in non-magnetic Taylor-Couette flows.

Returning finally to our original motivation in terms of astrophysical accretion disks, we note that at least some aspects of the nonlinear equilibration must be quite different there from what is seen in Figure \ref{profiles}. In particular, in accretion disks the angular velocity profile cannot possibly deviate as strongly from the basic Keplerian profile; if the mass of the central object is dominant, its gravity will always enforce a profile extremely close to $\Omega\sim r^{-3/2}$. This motivated us to perform an additional calculation in which the azimuthally and axially averaged flow profile was forced to remain identical to the laminar quasi-Keplerian profile, by adding a suitable body force in the Navier-Stokes equation which always drives these components of $\bf U$ back to the desired flow profile. (Numerically this is done by simply not time-stepping those components of $\bf U$.) The dynamo continued to exist in this calculation, and even has levels of Maxwell and Reynolds stresses broadly similar to the previous ones, as seen in Figure \ref{stresses}b. (For such a forced flow there are also stresses arising from the additional body force, but these do not affect the relative Maxwell, Reynolds and viscous contributions.)

These forced-flow results suggest that the dynamo presented here is not specific to Taylor-Couette flows, but is instead generic to any Rayleigh-stable differential rotation flows, including those in accretion disks. There are of course other important differences between Taylor-Couette geometry and accretion disks, including the boundaries in both radial and axial directions, stratification, compressibility, etc.\ \cite{Balbus1}. Nevertheless, the generic existence of such a dynamo in Rayleigh-stable flows suggests that accretion disks can operate with self-sustained magnetic fields, without relying on fields emanating from the central object (contrary to the shearing-box results of \cite{Pessah}, who suggested that angular momentum transport is negligible in the absence of externally imposed axial fields). Further work will map out the full range in parameter space where this dynamo exists, and quantify how the angular momentum transport varies with $Re$ and $Rm$, as well as the axial length $L_z$, which has been shown to be an important parameter in shearing-box simulations \cite{Nauman}.

This work was supported by the German Research Foundation (DFG) grant AV 120/1-1, and the National Science Foundation grant NSF PHY-1125915. Computing time from the North-German Supercomputing Alliance (HLRN) is also gratefully acknowledged.


\begin{thebibliography}{}
\bibitem{Roberts}
P. H. Roberts and E. M. King, Rep.\ Prog.\ Phys.\ 76, 096801 (2013).
\bibitem{Charbonneau}
P. Charbonneau, Ann.\ Rev.\ Astron.\ Astrophys.\ 52, 251 (2014).
\bibitem{Widrow}
L. M. Widrow, Rev.\ Mod.\ Phys.\ 74, 775 (2002).
\bibitem{Brandenburg1}
A. Brandenburg and K. Subramanian, Phys.\ Rep.\ 417, 1 (2005).
\bibitem{Balbus1}
S. A. Balbus, Annu.\ Rev.\ Astron.\ Astrophys.\ 41, 555 (2003).
\bibitem{Rayleigh}
Lord Rayleigh, Proc.\ R. Soc.\ A 93, 148 (1917).
\bibitem{Ji1}
H. Ji, M. Burin, E. Schartman and J. Goodman, Nature 444, 343 (2006).
\bibitem{Edlund}
E. M. Edlund and H. Ji, Phys.\ Rev.\ E 89, 021004 (2014).
\bibitem{Lesur}
G. Lesur and P. Y. Longaretti, Astron.\  Astrophys.\ 444, 25 (2005).
\bibitem{Avila}
M. Avila, Phys.\ Rev.\ Lett.\ 108, 124501 (2012).
\bibitem{Ostilla}
R. Ostilla-M\'onico, R. Verzicco, S. Grossman, and D. Lohse, J. Fluid Mech.\ 748, R3 (2014).
\bibitem{Lopez}
J. M. Lopez and M. Avila, J. Fluid Mech.\ 817, 21 (2017).
\bibitem{Shi}
L. Shi, B. Hof, M. Rampp and M. Avila, Phys.\ Fluids 29, 044107 (2017).
\bibitem{Velikhov}
E. P. Velikhov, Sov.\ Phys.\ JETP 36, 995 (1959).
\bibitem{Balbus2}
S. A. Balbus and J. F. Hawley, Astrophys.\ J. 376, 214 (1991).
\bibitem{Rudiger}
G. R\"udiger and Y. Zhang, Astron.\ Astrophys.\ 378, 302 (2001).
\bibitem{Ji2}
H. T. Ji, J. Goodman, and A. Kageyama, Mon.\ Not.\ R. Astron.\ Soc.\ 325,
L1 (2001)
\bibitem{Hollerbach1}
R. Hollerbach and G. R\"udiger, Phys.\ Rev.\ Lett.\ 95, 124501 (2005).
\bibitem{Stefani}
F. Stefani, T. Gundrum, G. Gerbeth, G. R\"udiger, M. Schultz, J. Szklarski
and R. Hollerbach, Phys.\ Rev.\ Lett.\ 97, 184502 (2006).
\bibitem{Flanagan}
K. Flanagan, M. Clark, C. Collins, C. M. Cooper, I. V. Khalzov, J. Wallace and
C. B. Forest, J.\ Plasma Phys.\ 81, 345810401 (2015).
\bibitem{Christensen}
U. Christensen, P. Olson and G. A. Glatzmaier, Geophys.\ J. Int.\ 138, 393
(1999).
\bibitem{Ponty}
Y. Ponty, J.-P. Laval, B. Dubrulle, F. Daviaud and J.-F. Pinton,
Phys.\ Rev.\ Lett.\ 99, 224501 (2007).
\bibitem{Krstulovic}
G. Krstulovic, G. Thorner, J.-P. Vest, S. Fauve and M. Brachet,
Phys.\ Rev.\ E 84, 066318 (2011).
\bibitem{Brandenburg2}
A. Brandenburg, A. Nordlund, R. F. Stein and U. Torkelsson, Astrophys.\ J.
446, 741 (1995).
\bibitem{Hawley}
J. F. Hawley, C. F. Gammie and S. A. Balbus, Astrophys.\ J. 464, 690 (1996).
\bibitem{Fromang}
S. Fromang, J. Papaloizou, G. Lesur and T. Heinemann, Astron.\ Astrophys.\ 476,
1123 (2007).
\bibitem{Yousef}
T. A. Yousef, T. Heinemann, F. Rincon, A. A. Schekochihin, N. Kleeorin,
I. Rogachevskii, S. C. Cowley and J. C. McWilliams, Astron.\ Nachr.\ 329,
737 (2008).
\bibitem{Johansen}
A. Johansen and Y. Levin, Astron.\ Astrophys.\ 490, 501 (2008).
\bibitem{Riols}
A. Riols, F. Rincon, C. Cossu, G. Lesur, P.-Y. Longaretti, G. I. Ogilvie
and J. Herault, J.\ Fluid Mech.\ 731, 1 (2013).
\bibitem{Kunz}
M. W. Kunz, J. M. Stone and E. Quataert, Phys.\ Rev.\ Lett.\ 117, 235101 (2016).
\bibitem{Nauman}
F. Nauman and M. E. Pessah, Astrophys.\ J. 833, 187 (2016).
\bibitem{Willis}
A. P. Willis and C. F. Barenghi, Astron.\ Astrophys.\ 393, 339 (2002).
\bibitem{Nore1}
C. Nore, J.-L. Guermond, R. Laguerre and J. L\'eorat, Phys.\ Fluids 24, 094106
(2012).
\bibitem{Gissinger}
C. Gissinger, Phys.\ Fluids 26, 044101 (2014).
\bibitem{Nore2}
C. Nore, D. Castanon Quiroz, L. Cappanera and J.-L. Guermond, EPL 114, 65002
(2016).
\bibitem{Gailitis}
A. Gailitis, O. Lielausis, S. Dement'ev, E. Platacis, A. Cifersons, G. Gerbeth, 
T. Gundrum, F. Stefani, M. Christen, H. H\"anel and G. Will, Phys.\ Rev.\ Lett.\
84, 4365 (2000).
\bibitem{Monchaux}
R. Monchaux, M. Berhanu, M. Bourgoin, M. Moulin, P. Odier, J.-F. Pinton, R. Volk, 
S. Fauve, N. Mordant, F. P\'etr\'elis, A. Chiffaudel, F. Daviaud, B. Dubrulle, 
C. Gasquet, L. Mari\'e and F. Ravelet, Phys.\ Rev.\ Lett.\ 98, 044502 (2007).
\bibitem{Guseva1}
A. Guseva, A. P. Willis, R. Hollerbach and M. Avila, New J. Phys.\ 17, 093018
(2015).
\bibitem{Guseva2}
A. Guseva, R. Hollerbach, A. P. Willis and M. Avila, Magnetohydrodynamics,
53, 25 (2017).
\bibitem{Guseva3}
A. Guseva, A. P. Willis, R. Hollerbach and M. Avila, arXiv:1705.03785
\bibitem{Hollerbach2}
R. Hollerbach, V. Teeluck and G. R\"udiger, Phys.\ Rev.\ Lett.\ 104, 044502
(2010).
\bibitem{Seilmayer}
M. Seilmayer, V. Galindo, G. Gerbeth, T. Gundrum, F. Stefani, M. Gellert,
G. R\"udiger, M. Schultz and R. Hollerbach, Phys.\ Rev.\ Lett.\ 113, 024505
(2014).
\bibitem{Dormy}
{\it Mathematical Aspects of Natural Dynamos}, edited by E. Dormy and A. M.
Soward (CRC Press, London, 2007).
\bibitem{Brauckmann}
H. J. Brauckmann and B. Eckhardt, J. Fluid Mech.\ 815, 149 (2017).
\bibitem{Pessah}
M. E. Pessah, C. Chan and D. Psaltis, Astrophys.\ J. 668, L51 (2007).
\end{thebibliography}
\end{document}